\documentclass[aps,amsfonts,pra,twocolumn,showpacs,floatfix]{revtex4}
\usepackage{epsfig,amsmath,amssymb,bm,epsf,graphics}

\def\ket#1{\vert#1\rangle}
\def\ketbra#1{\vert#1\rangle\langle#1\vert}
\def\ipr#1#2{\langle#1\vert#2\rangle}

\def\Longarrow{\protect\@lra}
\def\@lra{\relbar\joinrel\relbar\joinrel\relbar\joinrel%
          \relbar\joinrel\rightarrow}

\def\coe#1{{({\rm co}E_{#1})}}
\begin{document}
\title{Application of the geometric measure 
of entanglement to three-qubit mixed states}

\author{Tzu-Chieh Wei and Paul M. Goldbart}
\affiliation{Department of Physics, 
University of Illinois at Urbana-Champaign, 
1110 West Green Street, Urbana, Illinois 61801-3080, U.S.A.}

\date{March 26, 2003}

\begin{abstract}
The geometric measure of entanglement, originated by Shimony and 
by Barnum and Linden, is determined for a family of tripartite 
mixed states consisting of aribitrary mixtures of GHZ, W, and 
inverted-W states.  For this family of states, other measures 
of entanglement, such as entanglement of formation and relative 
entropy of entanglement, have not been determined.  The results 
for the geometric measure of entanglement are compared with those 
for negativity, which are also determined.  The results for the 
geometric measure of entanglement and the negativity provide 
examples of the determination of entanglement for nontrivial 
mixed multipartite states. 
\end{abstract}
\maketitle

\noindent
{\it Introduction\/}:   
In a recent Paper~\cite{WeiGoldbart02}, we have explored the issue 
of quantifying entanglement by invoking certain simple elements of 
Hilbert space geometry.  In doing this, we have been following a path 
originated by Shimony~\cite{Shimony95} and by Barnum and 
Linden~\cite{BarnumLinden01}.  Evidence for the reasonableness 
of this measure has been obtained via the investigation of 
certain bipartite mixed states for which other measures of 
entanglement (such as entanglement of formation and negativity) 
can also be computed.

One of the virtues of the geometric approach to entanglement is its 
straightforward adaptability to any arbitrary multipartite systems 
(of finite dimension).  In Ref.~\cite{WeiGoldbart02}, we described 
a procedure for applying the geometric measure of entanglement (GME)
to certain highly symmetric multipartites states, and illustrated 
this procedure via various examples.  The aim of the present Paper 
is to present the analytical computation  of the geometric measure 
of entanglement for a family of tripartite mixed states consisting 
of aribitrary mixtures of GHZ, W, and inverted-W states (these pure 
states being defined below).  For this family, other measures of 
entanglement---such as entanglement of formation and relative 
entropy of entanglement---have not been computed analytically.  (An 
exception is the measure of entanglement known as negativity, which, 
as we shall see, is readily computable.)

We are motivated to study the quantification of entanglement in 
multipartite mixed states for the basic reason that entanglement 
has been identified as a resource central to much of quantum 
information processing~\cite{NielsenChuang00}.  To date, progress 
in the quantification of entanglement for mixed states has resided 
primarily in the domain of bipartite systems~\cite{Horodecki01}.  
For multipartite systems in pure and mixed states the quantification
of entanglement presents even greater challenges.  

\smallskip
\noindent
{\it Basic geometric ideas; formulation of GME\/}:  
We begin by  briefly reviewing the formulation of this geometric 
measure in both pure-state and mixed-state settings.  
(For a discussion of the physical meanings of this measure, 
see Ref.~\cite{WeiGoldbart03}.)\thinspace\ 
Let us start with a multipartite system comprising $n$ parts, 
each of which can have a distinct Hilbert space.  Consider a 
general, $n$-partite, pure state (expanded in the local bases 
$\{|e_{p_i}^{(i)}\}$): 
$|\psi\rangle=\sum_{p_1\cdots p_n}\chi_{p_1p_2\cdots p_n}
|e_{p_1}^{(1)}e_{p_2}^{(2)}\cdots e_{p_n}^{(n)}\rangle$.
As shown in \cite{WeiGoldbart02}, the closest separable pure state 
\begin{equation}
\ket{\phi}\equiv\otimes_{i=1}^n|\phi^{(i)}\rangle=\otimes_{i=1}^{n}
\Big(\sum_{p_i}c_{p_i}^{(i)}\,|e_{p_i}^{(i)}\rangle\Big),
\end{equation}
satisfies the stationarity conditions
\begin{subequations}
\label{eqn:Eigen}
\begin{eqnarray}
\!\!\!\!\!\!\!\sum_{p_1\cdots\widehat{p_i}\cdots p_n}
\chi_{p_1p_2\cdots p_n}^*c_{p_1}^{(1)}\cdots\widehat{c_{p_i}^{(i)}}\cdots c_{p_n}^{(n)}=
\Lambda\,{c_{p_i}^{(i)}}^*, \\ 
\!\!\!\!\!\!\!\!\!\!\sum_{p_1\cdots\widehat{p_i}\cdots p_n}\chi_{p_1p_2\cdots p_n} {c_{p_1}^{(1)}}^*\cdots\widehat{{c_{p_i}^{(i)}}^*}\cdots {c_{p_n}^{(n)}}^*=
\Lambda\,c_{p_i}^{(i)}\,,
\end{eqnarray}
\end{subequations} 
in which the eigenvalue $\Lambda\in[-1,1]$ is associated with the Lagrange 
multiplier enforcing the constraint 
$\ipr{\phi}{\phi}\!=\!1$, 
and the symbol \,\,$\widehat{}$\,\, denotes exclusion.  
Moreover, the spectrum  $\Lambda$'s can be interpreted as the cosine of the
angle between $|\psi\rangle$ and $\ket{\phi}$; 
the largest, $\Lambda_{\max}$, which we call the 
{\it entanglement eigenvalue\/}, corresponds to the closest 
separable state. We shall adopt $E_{\sin^2}\equiv 1-\Lambda_{\max}^2$ as our
entanglement measure for any pure state $|\psi\rangle$; we shall, in the following, drop the subscript \lq$\max$\rq. 

Given the definition of entanglement for pure states just formulated, 
the extension to mixed states $\rho$ can be built upon pure states 
and is made via the use of the 
{\it convex hull\/} construction (indicated by ``co''), 
as was done for the entanglement of formation 
(see, e.g., Ref.~\cite{Wootters98}).  The essence is a minimization 
over all decompositions $\rho=\sum_i p_i\,|\psi_i\rangle\langle\psi_i|$ 
into pure states: 
\begin{eqnarray}
\label{eqn:Emixed}
E(\rho)
\equiv
\coe{\sin^2}(\rho)
\equiv
{\min_{\{p_i,\psi_i\}}}
\sum\nolimits_i p_i \, 
E_{\sin^2}(|\psi_i\rangle).
\end{eqnarray}
This convex hull construction ensures that the measure gives zero 
for separable states; however, in general it also complicates the 
task of determining mixed state entanglement.  As mentioned in the 
Introduction, the principal aim of the present Paper is to calculate 
the GME for a specific, nontrivial class of tripartite mixed states. 

\smallskip
\noindent
{\it Arbitrary mixture of GHZ, W, and inverted-W states\/}: 
Our goal is to calculate the GME for states of the form
\begin{equation}
\label{eqn:GWW}
\rho(x,y)\equiv x\ketbra{GHZ}+y\ketbra{W}+(1-x-y)\ketbra{\widetilde{W}},
\end{equation}
where $x,y\ge 0$ and $x+y\le 1$.  
The three relevant pure states are defined via 
\begin{subequations}
\begin{eqnarray}
\ket{GHZ}&\equiv&(\ket{000}+\ket{111})/{\sqrt{2}},\\
\ket{W}&\equiv&(\ket{001}+\ket{010}+\ket{100})/{\sqrt{3}},\\
\ket{\widetilde{W}}&\equiv&(\ket{110}+\ket{101}+\ket{011})/{\sqrt{3}}, 
\end{eqnarray}
\end{subequations}
and have the following basic features. 
For $\ket{GHZ}$, $\ket{000}$ and $\ket{111}$ are the closest
separable states, and for it $E_{\sin^2}(GHZ)=1/2$.  
For the W [or inverted-W] state,
$(\sqrt{2/3}\ket{0}+\sqrt{1/3}\ket{1})^{\otimes 3}$ 
[or $(\sqrt{2/3}\ket{1}+\sqrt{1/3}\ket{0})^{\otimes 3}$] 
is one of the closest separable states, and 
$E_{\sin^2}(W)=E_{\sin^2}(\widetilde{W})=5/9 >E_{\sin^2}(GHZ)$.
\begin{figure}[t]
\centerline{\psfig{figure=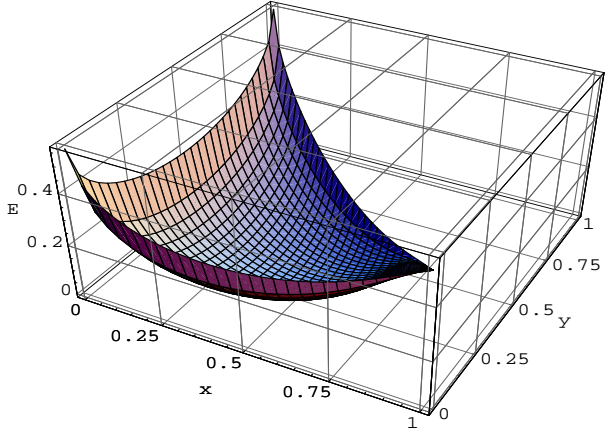,width=8cm,height=6cm,angle=0}}
\vspace{- 0.4cm}
\caption{Entanglement vs.~the composition of the pure state 
$\ket{\psi(x,y)}$.}
\label{fig:gGWW0}
\vspace{-0.2cm}
\end{figure}

A further property of the mixed state $\rho(x,y)$, which facilitates 
the computation of its entanglement, is a certain invariance, which 
we now describe.  Consider the local unitary transformation a single 
qubit: 
\begin{subequations}
\begin{eqnarray}
\ket{0}\rightarrow \ket{0}, \\
\ket{1}\rightarrow g^k\ket{1}, 
\end{eqnarray}
\end{subequations}
where $g=\exp{(2\pi i/3)}$, i.e., a relative phase shift.  
This transformation, when applied simultaneously to all 
three qubits, is denoted by $U_k$.  It is straightforward 
to see that $\rho(x,y)$ is invariant under the mapping 
\begin{equation}
P:\rho\rightarrow 
\frac{1}{3}\sum_{k=1}^3
U_k\,\rho\,U_k^\dagger\,.
\end{equation}
Thus, we can apply Vollbrecht-Werner 
technique~\cite{VollbrechtWerner01,WeiGoldbart02} 
to the compution of the entanglement of $\rho(x,y)$.
 
The first step of this task is to find the general form of the 
pure states that, under $P$, are projected to $\rho(x,y)$.  
This is readily seen to be 
\begin{equation}
\sqrt{x}\,e^{i\phi_1}\ket{GHZ}+
\sqrt{y}\,e^{i\phi_2}\ket{W}+
\sqrt{1-x-y}\,e^{i\phi_3}\ket{\widetilde{W}}.
\end{equation}
Of these, the least entangled state for given $(x,y)$ has all 
coefficients non-negative (up to a global phase), i.e., 
\begin{equation}
\ket{\psi(x,y)}\equiv
\sqrt{x}\ket{GHZ}+
\sqrt{y}\ket{W}+
\sqrt{1-x-y}\ket{\widetilde{W}}.
\end{equation}
The entanglement eigenvalue of $\ket{\psi(x,y)}$ can then be readily 
calculated (see Ref.~\cite{WeiGoldbart02} for the strategy), and one obtains
\begin{equation}
\Lambda(x,y)=\frac{1}{(1 \!+\! t^2)^{\frac{3}{2}}} \Big(\sqrt{\frac{x}{2}}(1 + t^3) + \sqrt{3y}\,t + 
      \sqrt{3(1\!-\!x\!-\!y)}\,t^2\Big), 
\end{equation}
where $t$ is the (unique) non-negative real root of the following 
third-order polynomial equation:
\begin{eqnarray}
&&3\sqrt{\frac{x}{2}}(-t + t^2) + \sqrt{3y}(-2t^2 + 1)
\nonumber\\ 
&&\qquad\qquad\quad
+\sqrt{3(1 - x-y)}(-t^3 + 2t) = 0.
\end{eqnarray}
Hence, the entanglement function for $\ket{\psi(x,y)}$, i.e., 
$E_\psi(x,y)\equiv 1-\Lambda(x,y)^2$, 
is determined (up to root-finding).
\begin{figure}[t]
\centerline{\psfig{figure=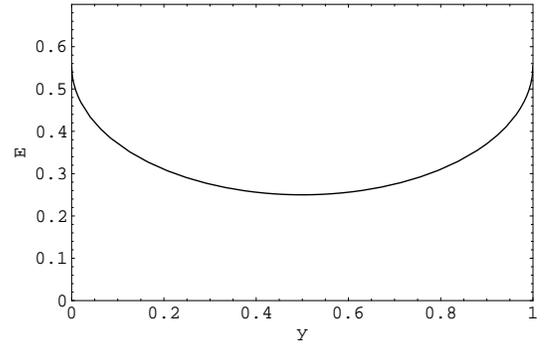,width=7cm,height=4.5cm,angle=0}}
\vspace{- 0.4cm}
\caption{Entanglement of the pure state
$\ket{\psi(x=0,y)}=
\sqrt{y}\,|{\rm W}\rangle+\sqrt{1-y}\,|\widetilde{\rm W}\rangle$ 
vs.~$y$.  This shows the entanglement along the $y$ axis.}  
\label{fig:WW}
\vspace{-0.4cm}
\end{figure}
\begin{figure}[h]
\centerline{\psfig{figure=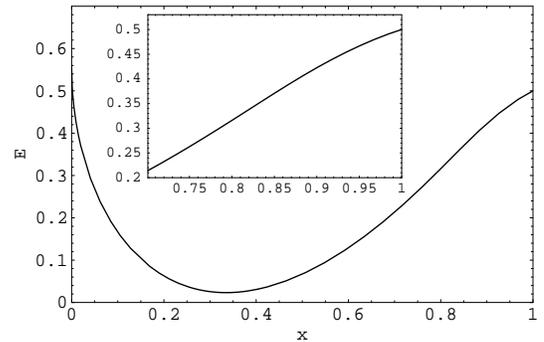,width=7cm,height=4.5cm,angle=0}}
\vspace{- 0.4cm}
\caption{Entanglement of the pure state 
$\ket{\psi(x,y=1-x)}=\sqrt{x}\,\ket{GHZ}+\sqrt{1-x}\,\ket{W}$ 
vs.~$x$.
This shows the entanglement along the diagonal boundary $x+y=1$.  
Note the absence of convexity near $x=1$; this region is 
repeated in the inset.}  
\label{fig:WGHZ}
\end{figure}

Recall that our aim is to determine the entanglement of the mixed 
state $\rho(x,y)$.  As we now know the entanglement of the 
corresponding pure state $\ket{\psi(x,y)}$, we may accomplish our aim by 
invoking a result due to Vollbrecht and Werner~\cite{VollbrechtWerner01}, 
which immediately gives the entanglement of $\rho(x,y)$ in terms of that 
of $\ket{\psi(x,y)}$, via the convex hull construction: 
$E_\rho(x,y)=(coE_\psi)(x,y)$.  Said in words, the entanglement surface 
$z=E_{\rho}(x,y)$ is the convex surface constructed from the surface 
$z=E_{\psi}(x,y)$.  

The idea underlying the use of the convex hull this.  Due to its 
linearity in $x$ and $y$, the state $\rho$ of Eq.~(\ref{eqn:GWW}) 
can [except when $(x,y)$ lies on the boundary] be decomposed into 
two parts: 
$\rho(x,y)=p \,\rho(x_1,y_1) +(1-p)\rho(x_2,y_2)$ 
with the weight $p$ and end-points $(x_1,y_1)$ and $(x_2,y_2)$ 
related by 
$p\, x_1+(1-p)x_2=x$ and 
$p\, y_1+(1-p)y_2=y$ .
Now, if it should happen that 
$p   E_{\psi}(x_1,y_1)+
(1-p)E_{\psi}(x_2,y_2)< 
     E_{\psi}(x,y)$ 
then the entanglement averaged over the end-points gives a value 
lower than that at the interior point $(x,y)$; this conforms with 
the convex-hull construction.

It should be pointed out that the convex hull should be taken 
with respect to parameters on which the density matrix depends 
{\it linearly\/}, as $x$ and $y$ do in the example of $\rho(x,y)$.  
Furthermore, in order to obtain the convex hull of a function, 
one needs to know the {\it global\/} structure of the function; 
in the present case, $E_{\psi}(x,y)$.  We note that numerical 
algorithms have been developed for constructing convex 
hulls~\cite{QHull}. 
\begin{figure}[t]
\centerline{\psfig{figure=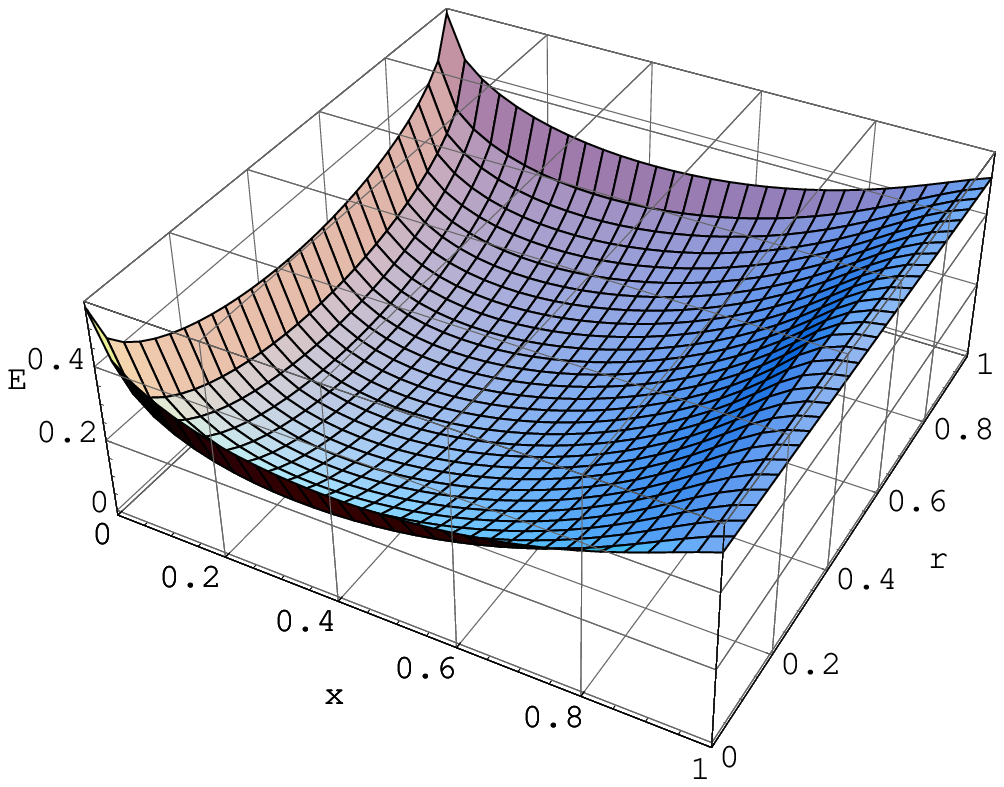,width=8cm,height=5.5cm,angle=0}}
\vspace{- 0.4cm}
\caption{Entanglement of the pure state 
$\ket{\psi\big(x,(1-x)r\big)}=\sqrt{x}\,
\ket{GHZ}+
\sqrt{(1-x)r}\,\ket{W}+
\sqrt{(1-x)(1-r)}\ket{\widetilde{W}}$ vs.~$x$ and $r$.  
Note the symmetry of the surface with respect with $r=1/2$.}  
\label{fig:gGWWxr0}
\vspace{-0.2cm}
\end{figure}
\begin{figure}[t]
\centerline{\psfig{figure=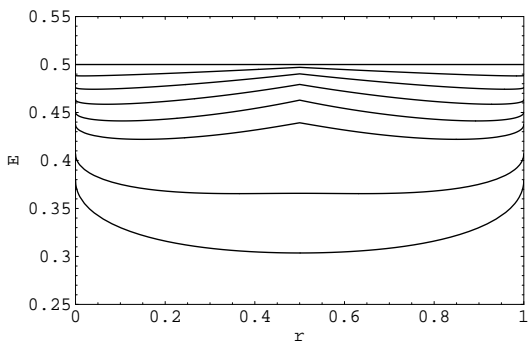,width=7cm,height=4.5cm,angle=0}}
\vspace{- 0.4cm}
\caption{Entanglement of the pure states
$\ket{\psi\big(x,(1-x)r\big)}=
\sqrt{x}\,
\ket{GHZ}+
\sqrt{(1-x)r}\,\ket{W}+
\sqrt{(1-x)(1-r)}\ket{\widetilde{W}}$
vs.~$r$ for various values of $x$ 
(from the bottom: 0.8, 0.85, 0.9, 0.92, 0.94, 0.96, 0.98, 1). 
This reveals the nonconvexity in $r$ for intermediate values of $x$.}  
\label{fig:gGWWxr1}
\end{figure}
\begin{figure}[t]
\centerline{\psfig{figure=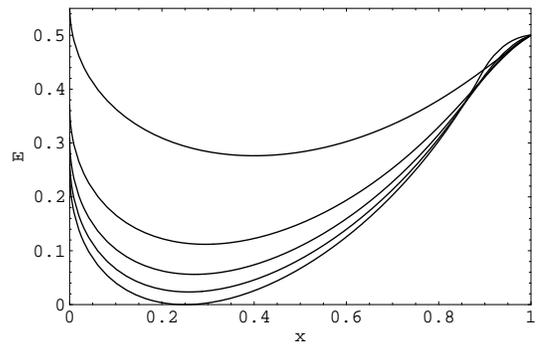,width=7cm,height=4.5cm,angle=0}}
\vspace{- 0.2cm}
\caption{Entanglement of the pure states 
$\ket{\psi\big(x,(1-x)r\big)}=
\sqrt{x}\,\ket{GHZ}+
\sqrt{(1-x)r}\,\ket{W}+
\sqrt{(1-x)(1-r)}\ket{\widetilde{W}}$ vs.~$x$ 
for various values of $r$ 
(from the top: 0, 0.1, 0.2, 0.3, 0.5). 
This reveals the nonconvexity in $x$ in the (approximate) 
interval $[0.85,1]$.}  
\label{fig:gGWWxr2}
\vspace{-0.4cm}
\end{figure}

As we have discussed, our route to establishing the entanglement of 
$\rho(x,y)$ invloves the analysis of the entanglement of $\ket{\psi(x,y)}$, 
which we show in Fig.~\ref{fig:gGWW0}.  Although it is not obvious, 
the corresponding surface fails to be convex near to the point $(x,y)=(1,0)$, 
and therefore, in this region, we must suitably convexify in order to 
obtain  the entanglement of $\rho(x,y)$.  To illustrate the properties 
of the entanglement of $\ket{\psi(x,y)}$ we show, in Fig.~\ref{fig:WW}, 
the entanglement of $\ket{\psi(x,y)}$ along the line $x=0$; 
evidently this is convex.  By contrast, along the line $x+y=1$, 
there is a region in which the entanglement is not convex, as 
Fig.~\ref{fig:WGHZ} shows.  The nonconvexity of the entanglement of 
$\ket{\psi(x,y)}$ complicates the calculation of the entanglement of 
$\rho(x,y)$, as it necessitates a procedure for constructing the 
convex hull in the (as it happens, small) nonconvex region.
Elsewhere in the $xy$ plane the entanglement of $\rho(x,y)$ is given 
directly by the entanglement of $\ket{\psi(x,y)}$.  

\begin{figure}[t]
\centerline{\psfig{figure=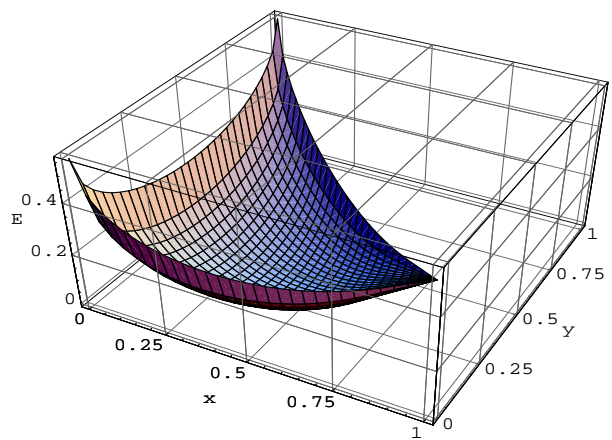,width=8cm,height=6cm,angle=0}}
\vspace{- 0.2cm}
\caption{Entanglement of the mixed state ${\rho(x,y)}$.}
\label{fig:gGWW}
\end{figure} 
At worst, convexification would have to be undertaken numerically.  
However, in the present setting it turns out one can determine the 
convex surface essentially analytically, by performing the necessary 
surgery on surface $z=E_{\psi}(x,y)$.  To do this, we make use of the 
fact that if we parametrize $y$ via $(1-x)r$, i.e., we consider 
\begin{eqnarray}
{\rho\big(x,(1-x)r\big)}&=&x\,
\ketbra{GHZ}+(1-x)r\,\ketbra{W}\nonumber \\
&&+(1-x)(1-r)\ketbra{\widetilde{W}}, 
\label{eqn:Rhoxr}
\end{eqnarray}
where $0\le r\le 1$ [and similarly for $\ket{\psi(x,y)}$] then, 
as a function of $(x,r)$, the entanglement will be symmetric with 
respect to $r=1/2$, as Fig.~\ref{fig:gGWWxr0} makes evident. 
With this parametrization, the nonconvex region of the entanglement 
of $\ket{\psi}$ can more clearly be identified.  

To convexify this surface we adopt the following convenient strategy. 
First, we reparametrize the coordinates, exchanging $y$ by $(1-x)r$.  
Now, owing to the linearity, in $r$ at fixed $x$ and vice versa, of the 
coefficients $x$, $(1-x)r$ and $(1-x)(1-r)$ in Eq.~(\ref{eqn:Rhoxr}), 
it is certainly necessary for the entanglement of $\rho$ to be a 
convex function of $r$ at fixed $x$ and vice versa.  Convexity is, 
however, not necessary in other directions in the $xr$ plane, owing to 
the nonlinearity of the the coefficients under simultaneous variations 
of $x$ and $r$.  Put more simply: convexity is not necessary throughout 
the $(x,r)$ plane because straight lines in the $(x,r)$ plane 
do not correspond to straight lines in the $(x,y)$ plane
(except along lines parallel either to the $r$ or the $x$ axis).
Thus, our strategy will be to convexify in a restricted sense: 
first along lines parallel to the $r$ axis and then along lines 
parallel to the $x$ axis.  Having done this, we shall check to see 
that no further convexification is necessary.

For each $x$, we convexify the curve $z=E_{\psi}\big(x,(1-x)r\big)$ as 
a function of $r$, and then generate a new surface by allowing $x$ to 
vary.  More specifically, the nonconvexity in this direction has the 
form of a symmetric pair of minima located on either side of a cusp, 
as shown in Fig.~\ref{fig:gGWWxr1}.  Thus, to correct for it, we 
simply locate the minima and connect them by a straight line. 

What remains is to consider the issue of convexity along the $x$ 
(i.e., at fixed $r$) direction for the surface just constructed.  
In this direction, nonconvexity occurs when $x$ is, roughly speaking, 
greater than $0.8$, as Fig.~\ref{fig:gGWWxr2} suggests.  In contrast 
with the case of nonconvexity at fixed $r$, this nonconvexity is due 
to an inflection point at which the second derivative vanishes.
To correct for it, we locate the point $x=x_0$ such that the tangent 
at $x=x_0$ is equal to that of the line between the point on the curve 
at $x_0$ and the end-point at $x=1$, and connect them with a straight 
line.  This furnishes us with a surface convexified with respect to 
$x$ (at fixed $r$) and vice versa. 

\begin{figure}[t]
\centerline{\psfig{figure=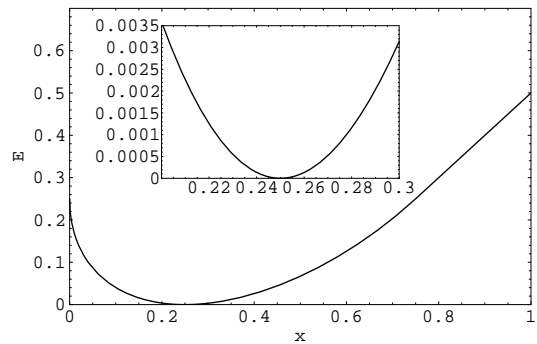,width=7cm,height=4.5cm,angle=0}}
\vspace{- 0.4cm}
\caption{Entanglement of the mixed state 
${\rho\big(x,y=(1-x)/2\big)}=x\,
\ketbra{GHZ}+
\frac{1-x}{2}\big(\ketbra{W}+
\ketbra{\widetilde{W}}\big)$ vs.~$x$.  
Inset: enlargement of the region $x\in[0.2,0.3]$. 
This contains the only point, $(x,y)=(1/4,3/8)$, 
at which $E_{\rho}(x,y)$ vanishes.}  
\label{fig:gGWW05}
\vspace{-0.4cm}
\end{figure}

Armed with this surface, we return to the $(x,y)$ parametrization, 
and ask whether or not it is fully convex (i.e., convex along straight 
lines connecting {\it any\/} pair of points).  Said equivalently, we 
ask whether or not any further convexification is required.  Although 
we have not proven it, on the basis of extensive numerical exploration 
we are confident that the resulting surface is, indeed, convex.  The 
resulting convex entanglement surface for $\rho(x,y)$ is shown in 
Fig.~\ref{fig:gGWW}.  Figure~\ref{fig:gGWW05} exemplifies this 
convexity along the line $2y+x=1$.  We have observed that for the 
case at hand it is adequate to correct for nonconvexity only in the 
$x$ direction at fixed $r$.

\begin{figure}[t]
\centerline{\psfig{figure=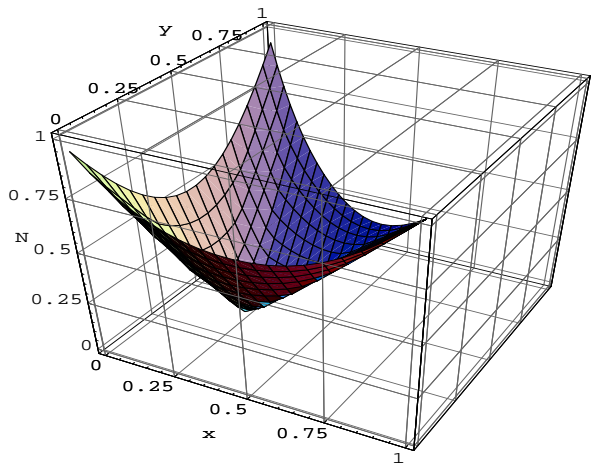,width=7.8cm, height=6.2cm,angle=0}}
\vspace{- 0.3cm}
\caption{Negativity of the mixed state ${\rho(x,y)}$.}
\label{fig:Nxy}
\vspace{-0.5cm}
\end{figure}

\smallskip
\noindent
{\it Negativity\/}: 
This measure of entanglement is defined as twice the absolute value 
of the sum of the negative eigenvalues of the partial transpose of 
the density
matrix~\cite{ZyczkowskiWerner,WeiNemotoGoldbartKwiatMunroVerstraete03}.
In the present setting, 
viz., the family $\rho(x,y)$ of three-qubit states, the partial 
transpose may equivalently be taken with respect to any one of the 
three parties, owing to the invariance of $\rho(x,y)$ under all 
permutations of the parties.  Transposing with respect to the third 
party, one has 
\begin{equation}
N(\rho)\equiv-2\sum_{\lambda_i<0} \lambda_i, 
\end{equation}
where the $\lambda$'s are the eigenvalues of the matrix $\rho^{T_3}$, 

It is straightforward to calculate the negativity for $\rho(x,y)$; 
the results are shown in Fig.~\ref{fig:Nxy}.  
It is interesting to note that, for all allowable
ranges of $(x,y)$, the state $\rho(x,y)$ has nonzero negativity, except
at $(x,y)=(1/4,3/8)$, at which the calculation of the GME shows that 
the density matrix is indeed separable.  The fact that the only 
positve-partial-transpose (PPT)
state is separable is the statement that there are no entangled PPT states 
(i.e., no PPT bound entangled states) within this family of three-qubit 
mixed states.  The negativity surface, Fig.~\ref{fig:Nxy}, is 
qualitatively---but not quantitatively---the same as that of GME. 
By inspecting the negativity and GME surfaces one can see that they 
present ordering difficulties.  We mean by this that one can find 
pairs of states $\rho(x_1,y_1)$ and $\rho(x_2,y_2)$ that respectively 
have negativities $N_1$ and $N_2$ and GMEs $E_1$ and $E_2$ such 
that, say, $N_1< N_2$ but $E_1>E_2$.  Said equivalently, the negativity 
and the GME do not necessarily agree on which of a pair of states is 
the more entangled.  For two qubit settings, such ordering difficulties 
do not show up for pure states but can for mixed states~\cite{Ordering,WeiNemotoGoldbartKwiatMunroVerstraete03}.  
On the other hand, such difficulties already show up for pure states, 
as the following example shows: $N(GHZ)=1>N(W)=2\sqrt{2}/3$ 
whereas for the GME the order is reversed.  We note that for the 
relative entropy of entanglement $E_R$, one has $E_R(GHZ)=\log 2 < E_R(W)=\log(9/4)$\cite{PlenioVedral01}.

\smallskip
\noindent
{\it Concluding remarks\/}: 
By making use of the geometric measure of entanglement we have addressed 
the entanglement of a rather general family of three-qubit mixed 
states analytically (up to root-finding).  This family consists 
of arbitrary mixtures of GHZ, W, and inverted-W states.  To the 
best of our knowledge, corresponding results have not, to date, 
been obtained for other measures of entanglement, such as 
entanglement of formation and relative entropy of entanglement.  
We have obtained  corresponding results for the 
negativity measure of entanglement, and have compared them with 
those for the geometric measure of entanglement.  Among other 
things, we have found that there are no PPT bound entangled states 
within this general family.

We are unaware of any explicit generalization of entanglement of 
formation to multipartite mixed states.  However, if such a generalization
should emerge, and if it should be based on the convex hull construction 
(as it is in the bipartite case), then one may be able to calculate the 
entanglement of formation for the family of mixed states considered 
in the present Paper.  It would, then, be interesting to know 
whether or not the similarities between entanglement of formation and the
geometric measure of entanglement
found at the level of certain bipartite mixed 
states~\cite{WeiGoldbart02} continue to hold beyond the bipartite 
world.  

\bigskip
\noindent
{\it Acknowledgments\/}: 
We thank J.~Altepeter, H.~Edelsbrunner, M.~Ericsson, P.~Kwiat, 
S.~Mukhopadhyay, F.~Verstraete and especially W.~J.~Munro for 
discussions.
This work was supported by 
NSF EIA01-21568. 
and DOE DEFG02-91ER45439.
TCW acknowledges a Mavis Memorial Fund Scholarship. 

\end{document}